# PorePy: An Open-Source Simulation Tool for Flow and Transport in Deformable Fractured Rocks

Eirik Keilegavlen[1], Alessio Fumagalli[1], Runar Berge[1], Ivar Stefansson[1], Inga Berre[1,2]
[1]Department of Mathematics, University of Bergen.
[2]Christian Michelsen Research
Corresponding author: Eirik.Keilegavlen@uib.no, Department of Mathematics, University of Bergen, Pb 7800, 5020 Bergen, Norway.

## Abstract

Fractures are ubiquitous in the subsurface and strongly affect flow and deformation. The physical shape of the fractures, they are long and thin objects, puts strong limitations on how the effect of this dynamics can be incorporated into standard reservoir simulation tools. This paper reports the development of an open-source software framework, termed PorePy, which is aimed at simulation of flow and transport in three-dimensional fractured reservoirs, as well as deformation of the reservoir due to shearing along fracture and fault planes. Starting from a description of fractures as polygons embedded in a 3D domain, PorePy provides semi-automatic gridding to construct a discrete-fracture-matrix model, which forms the basis for subsequent simulations. PorePy allows for flow and transport in all lower-dimensional objects, including planes (2D) representing fractures, and lines (1D) and points (0D), representing fracture intersections. Interaction between processes in neighboring domains of different dimension is implemented as a sequence of couplings of objects one dimension apart. This readily allows for handling of complex fracture geometries compared to capabilities of existing software. In addition to flow and transport, PorePy provides models for rock mechanics, poro-elasticity and coupling with fracture deformation models.

PorePy provides both finite-volume and virtual finite element discretizations. The code is implemented in *Python*, is easy to install and configure, and can be adapted and employed by means of high-level *Python* scripts. The software is fully open, and can serve as a framework for transparency and reproducibility of simulations. We describe the design principles of PorePy from a user perspective, with focus on possibilities within gridding, covered physical processes and available discretizations. The power of the framework is illustrated with two sets of simulations; involving respectively coupled flow and transport in a fractured porous medium, and low-pressure stimulation of a geothermal reservoir.

## 1. Introduction

Simulation of flow, transport and deformation of fractured rocks is of critical importance to several applications such as subsurface energy extraction and storage, and waste disposal. While the topics have received considerable attention the last decade, the development of reliable simulation tools remains a formidable challenge. Many reasons can be given for this deficiency, we here pinpoint three causes: First, while natural fractures are thin compared to the characteristic length of the domains of interest, their extension can span wide scales. Combined with the strongly heterogeneous flow properties of fractures, which can act both as conduits and barriers, this makes the implementation of a simulation tool that honors inter-dimensional couplings a highly challenging task. Second, the phenomena of practical interest tend to involve multi-physics couplings that are non-trivial for numerical simulations, such as interaction between flow, temperature evolution, geo-chemical effects and fracture deformation. Third, fracture networks have highly complex geometries, necessitating unstructured gridding and versatile discretization schemes.

Existing software packages that aims to incorporate flow-driven dynamics in fractured rock, and that have at least some degree of code accessibility, include the TOUGH2 family (Fakcharoenphol et al., 2013; Preuss, 1991), OpenGeoSys (Kolditz et al., 2012) and CSMP (Matthai et al., 2007). These frameworks focus on general-purpose, large scale simulations. Capabilities to simulate fractured media is also to some extent available in other open platforms, such as DuMuX (Flemisch et al., 2011), MRST (Lie et al., 2012) and OPM (Open Porous Media Team, 2017), but these offer no special treatment of fractures as such.

With models and simulation technology for dynamics in fractured rocks being continuously developed and improved, there is a need for a framework for rapid prototyping that readily can represent interaction between the physical processes and the fractured structure of the rock, with capabilities for exploration of different models and physical mechanisms. Moreover, with an eye on the increasing focus on transparency in research and reproducibility, full code accessibility is desirable. Here, we present what to our knowledge is the first open-source framework designed specifically for coupling flow, transport and deformation of fractured rocks. The software, termed PorePy, is implemented in *Python*, and can easily be adapted to multi-physics simulations using a high-level scripting language. In the following we describe the construction of mixed-dimensional grids that cover the rock matrix, fractures and fracture intersections. We further provide an overview of the physical processes included, and the corresponding available discretization schemes. To demonstrate the capacity and usage of the code, we show simulations of flow and transport in 3D fractured reservoirs, and of low-pressure stimulation of a fracture network, as applied in enhanced geothermal systems. The software is fully open source, see [www.github.com/pmgbergen/porepy](www.github.com/pmgbergen/porepy), and comes with an extensive set of examples and tutorials. The software is actively developed and maintained using modern development technology such as unit testing and continuous integration. Usage of the code is indicated by code snippets in this paper; the full source code of PorePy, together with scripts used for the simulations presented herein, can be found at the software webpage.

## 2. Simulation models for fractured reservoirs

Natural fracture networks are commonly characterized by complex intersection geometries. The key to a robust implementation of a simulation tool handling such networks is a systematic approach to the interaction between the different parts of the domain, consisting of i) the background three-dimensional (3D) porous medium, ii) the two-dimensional (2D) fracture surfaces, iii) the one-dimensional (1D) line intersections between two different fractures, and iv) the zero-dimensional (0D) point intersections between more than two different fractures. We refer to this as the mixed-dimensional geometrical features of the domain. The lower-dimensional features of the domain are important for modeling of flow and transport, but typically form an obstacle for meshing.

There is a rich literature on the representation of fractures in simulation models, for an overview confer (Aagaard et al., 2013; Berkowitz, 2002; Boon et al., 2016; Dietrich et al., 2005; Flemisch et al., 2018; Lee et al., 2001; Martin et al., 2005; Paluszny et al., 2007) and the references therein. A principal choice to be made is whether the fractures are explicitly represented, or represented by upscaled quantities. This decision involves several factors including the target application, the dynamic processes involved, the type and quality of data available, and the tradeoff between computational cost and accuracy. PorePy mainly targets applications where dynamics in fractures have a decisive impact, and therefore allows for explicit representation of main fractures, but with the option that some fractures are upscaled. More specifically, the simulation models in PorePy are based upon Discrete Fracture Matrix (DFM) modeling concepts (Berkowitz, 2002), wherein fractures are explicitly represented as objects of one dimension less than the matrix, with aperture as a parameter. Computationally, this is advantageous compared to equi-dimensional models as narrow grid cells are avoided. In addition, the approach is ideal for modeling of physics involving dynamic aperture change as in this case there is no need for re-meshing. Moreover, acknowledging the critical impact fracture intersections can have on flow (Peacock et al., 2017; Rotevatn et al., 2009), PorePy can, contrary to most DFM models, also handle dynamics along 1D intersections.

Although this paper mostly emphasizes three-dimensional problems, PorePy can also handle 2D domains containing 1D fracture lines. Except from the meshing, which is significantly simpler in two dimensions, the code is, in general, the same as for the 3D case.

## 2.1 Fractures and fracture networks

In PorePy, a fracture is represented as a polygon embedded in 3D space. For simplicity, PorePy considers only planar fractures, and it is assumed that all fracture polygons are convex. Fractures are specified either directly by their vertexes, or as ellipsoids that are converted to polygons, see Figure 1c for a representative code snippet.

The union of individual fractures form a fracture network, which is represented by as a separate object in PorePy. PorePy provides functionality for automatic identification of the lower-dimensional intersections between fractures in the network, with a specified tolerance to reflect data accuracy as well as finite accuracy in the computations. Figure 2 illustrates some types of intersections that can be handled.

## 2.2 Meshing of mixed-dimensional geometries

While the DFM approach is well suited to resolve the strong heterogeneities in parameters and dynamics between fractures and the surrounding matrix, most discretization schemes within the DFM family require a computational mesh that conforms to the fractures as well as their intersections. PorePy provides such computational grids based on a specification of the individual fractures, and relies on Gmsh (Geuzaine and Remacle, 2009) for the actual meshing. The meshing produces a 3D grid of the rock matrix, as well as 2D, 1D and 0D grids that represent fractures, and two levels of intersections, respectively, see Figure 1b-c for an illustration. Two-dimensional domains are handled in a similar manner. The procedure is automatic, with no need for the user to identify fracture intersections, or interacting with the meshing software, as indicated in Figure 1d. If the fracture geometry has features that are of the size of the tolerance, it may be necessary to either tweak the fracture network, or adjust the tolerance.

For some applications, e.g. permeability upscaling and transport in hard rocks, it may be of interest to ignore the rock matrix, and instead consider a discrete fracture network (DFN) model. This is easily achieved in PorePy by simply not meshing the 3D domain. The resulting mixed-dimensional grid, considering only fracture planes and intersection lines and points, can be treated by the same numerical schemes. Example usage of these features can be found in (Fumagalli and Keilegavlen, 2018).

a)
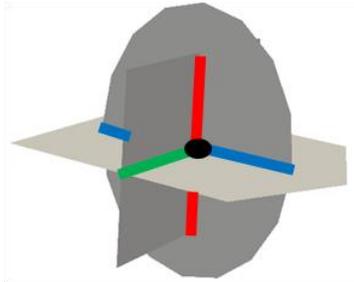

b)
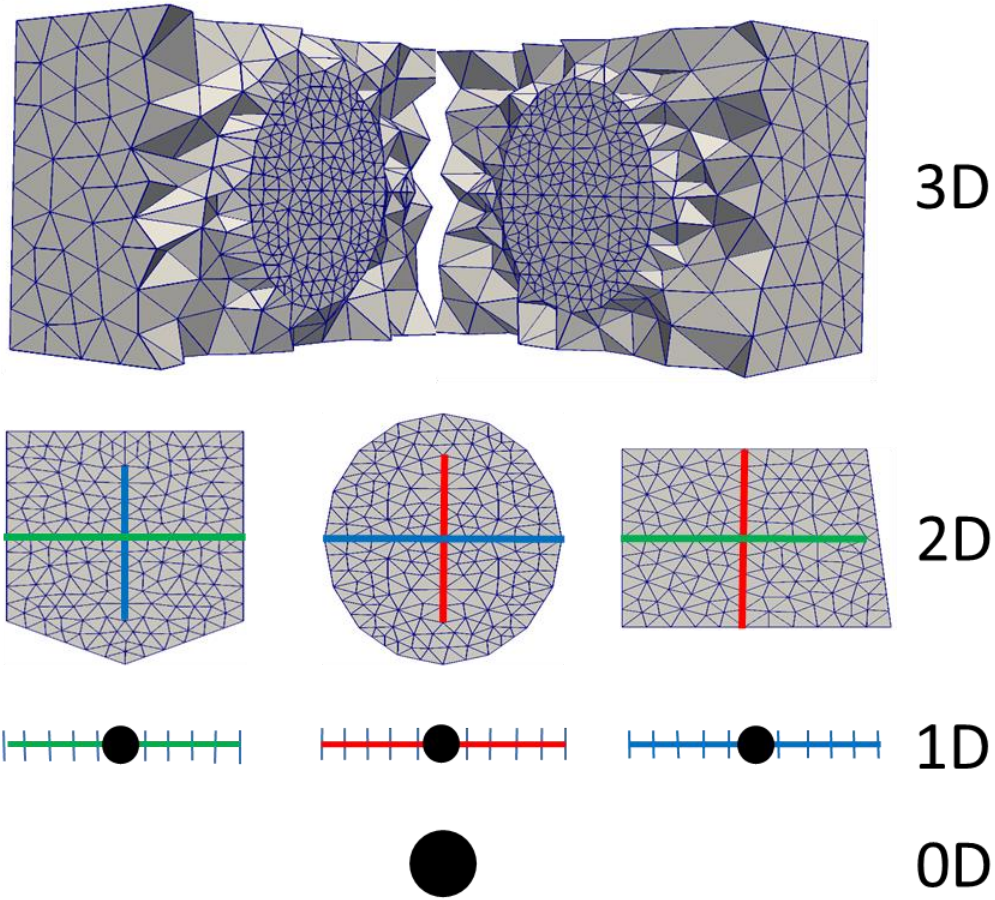

3D

2D

1D

0D

c)
```
f_1=Fracture([[-1,1.2,1,-1],[0,0,0,0],[-1,-1,1,1]])
f_2=EllipticFracture(center=[0, 0, 0],major_axis=1,  minor_axis=1,
         major_axis_angle=0, strike_angle=π/2, dip_angle=π/2)
f_3=Fracture([[-1,1,1.5,1,-1],[-1,-1,0,1,1],[0,0,0,0,0]])
mesh = meshing.simplex_grid([f_1, f_2, f_3])
```

*Figure 1: Conceptual figure for illustration of a fracture network, including meshing and lower-dimensional representation. a) Fracture network, b) Meshes in 3D, and subdimensions. Fracture intersections (1D) are represented by colored lines, the 0D grid by black circle. The 3D mesh is cut to expose the circular fracture. c) PorePy code used to define fractures and the mixed-dimensional grid.*

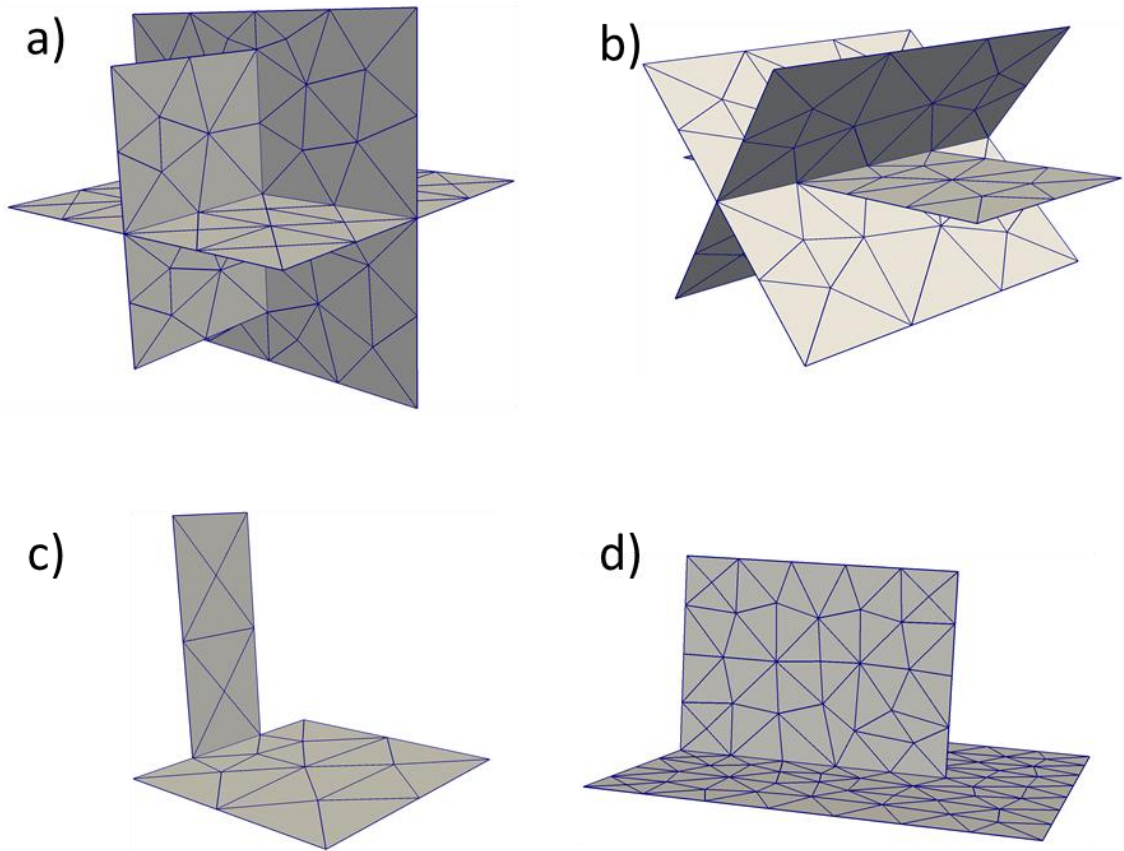

*Figure 2: Selection of intersection types handled by PorePy: a) Three intersections meeting in a point; b) fractures meeting in lines, with partly overlapping intersection lines; c) L-intersection; d) T-intersection.*

## 2.3 Import filters and mesh size control

In addition to specifying fractures by their vertexes, PorePy can also construct computational meshes directly from files containing fracture vertexes that may be output from a manual or automatic fracture tracing, e.g. (Hardebol and Bertotti, 2013). Figure 3 shows a fracture map representative for the result of fracture tracing, together with the two-dimensional computational mesh and the associated PorePy code. In this procedure, care is taken to preserve the identified topology in the geological characterization of the fracture network (Sanderson and Nixon, 2015).

The extension of outcrop data to a full 3D network is a challenging task, as discussed in (Bisdom et al., 2014; García-Sellés et al., 2011), which PorePy by itself is not designed to handle. However, functionality to extrude fracture traces based on a limited range of user specification is provided – extension to rules not covered must be handled on a case by case basis. As an example, Figure 4 shows how fractures from a 2D outcrop can be extended based on specified extrusion heights, and dip and strike angles, and then meshed. For realistic simulations, a more relevant approach may be the generation of 3D fractures based on statistics extracted from the outcrop.

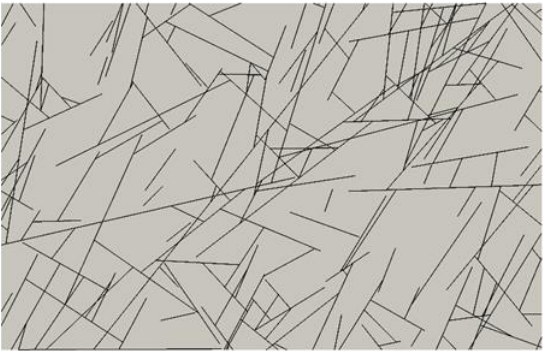
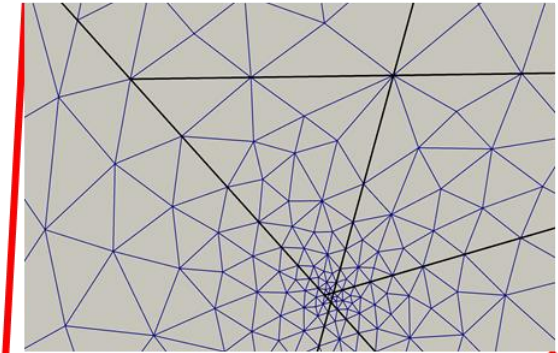
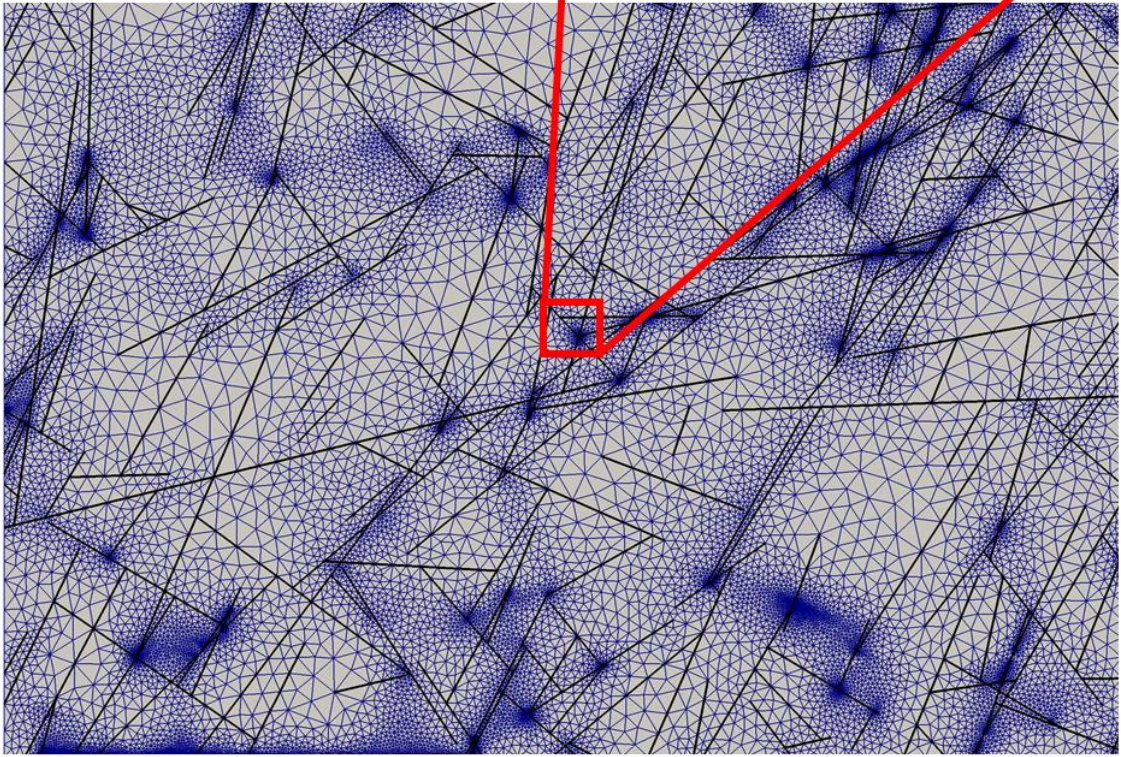

```
# Let mesh size adapt to neighboring geometric details
mesh_size = {'mode': 'weighted'}
# Read fracture information from file, generate mixed-dimensional mesh
mesh = importer.mesh_from_csv(file_name, mesh_size)
```

*Figure 3: Automatic meshing of a 2D domain: a) Fracture geometry, as specified by the user, b) the resulting mesh, c) zoom-in of b). As seen from the zoom-in, regions with close fracture intersections, or with small angles between fractures, have highly refined meshes.*

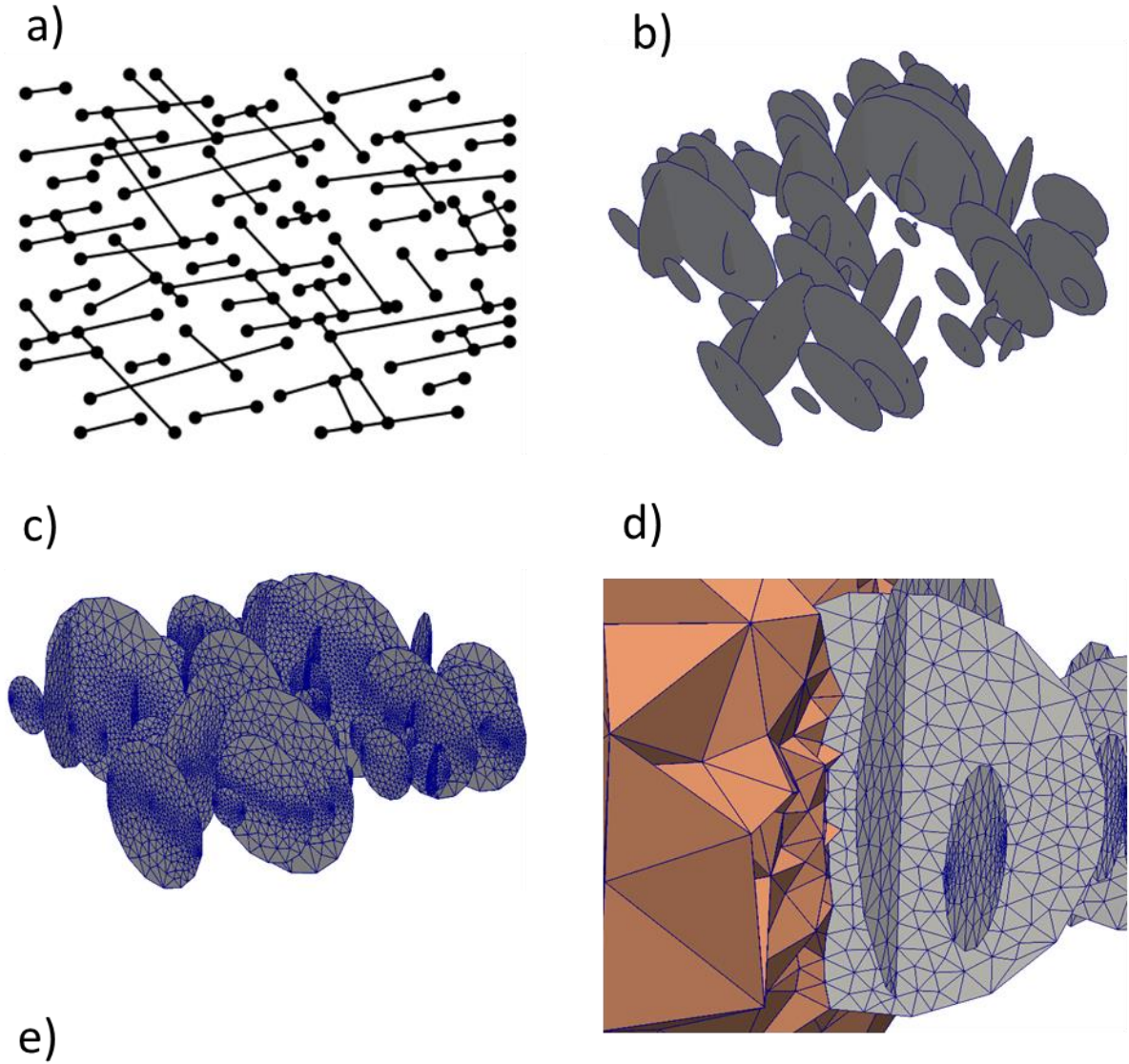

*Figure 4: Extrusion of 3D fracture network from outcrop trace and meshing. a) Hypothetical 2D outcrop containing 66 fractures; b) extrusion to consistent 3D fracture network; c) Generated mesh, showing 2D surfaces only; d) cut of 3D mesh, also indicating penetration by lower-dimensional grids; e) PorePy code needed to extrude fractures and generate mesh.*

While controlling the size and number of cells in the computational grid is key to balancing simulation cost and accuracy, it is also a technical exercise that the user ideally should be shielded from. This is particularly true for complex fracture networks, where details in the geometric configuration dictate the mesh size, at least locally. PorePy attempts to resolve this by automatically adapting the resolution to the local fracture geometry, and then let Gmsh coarsen the mesh towards a user specified background cell size, again, see Figure 3. However, as the fracture geometry is fixed, difficult geometric details such as small angles and short distances between geometrical objects may still cause poor quality and excessively small cells.

### 2.4 Data visualization

To visualize fracture networks, meshes and simulation results, PorePy provides export filters to Paraview (Ahrens et al., 2005). The export preserves the link between data and their associated objects (matrix, fractures, intersections).

## 3 Physical processes and discretizations

On top of the grid, PorePy provides simulation tools for three different physical processes: Fluid flow in the coupled mixed-dimensional matrix-fracture system, transport of a scalar (tracer or temperature) accounting for both advection and conduction, and the coupling of rock mechanics and displacement along fracture surfaces. Below, we give an overview of the core equations implemented in PorePy, together with the discretization schemes available. From this core, more elaborate models can rapidly be extended and applied to different simulation setups. Examples of this in the form of multi-physics couplings and interaction with properties of the fractures are provided in the next section.

PorePy will assign default values, typically of unit size, to all parameters needed to solve a given equation. These can be replaced by user defined values where needed. Spatially varying parameters, such as permeability, porosity and fracture aperture, are included as cell-based quantities in the discretization schemes discussed below. Thus, if desired, PorePy can accommodate spatial heterogeneities with a resolution dictated by the mesh size.

### 3.1 Fluid flow

We here focus on a model for a slightly compressible fluid; the simpler option of an incompressible fluid is also available, as indicated in Section 4.1.

#### 3.1.1 Model and parameters

Within the rock matrix, fractures and fracture intersections, the fluid flow rate $q$ is related to the gradient of the pressure $p$ via Darcy's law, $q = -K\nabla p$. For fractures and fracture intersections, only the components of the permeability $K$ laying in the plane and line, respectively, are considered. The fracture permeability is commonly related to the aperture by the cubic law (Berkowitz, 2002), thus both heterogeneities and changes in aperture are easily accommodated by simple programming from the user side. The flow between dimensions is modeled by an interface law (Martin et al., 2005)

$$\hat{q} \cdot \hat{n} = K_n(\hat{p} - \check{p}), \tag{1}$$

where $\hat{p}$ and $\check{p}$ are the pressures in the higher and lower-dimensional object, respectively, and $K_n$ is the transmissivity at the interface between dimensions. Conservation of mass (assuming a weakly compressible fluid) is expressed as

$$c_f \phi \frac{\partial p}{\partial t} + \nabla \cdot q = w, \tag{2}$$

where $c_f$ is the fluid compressibility, $\phi$ is porosity and $w$ represents source and sink terms, which may also include in and outflow from lower and higher dimensions.

Within PorePy, the parameters that need to be specified are permeability and porosity in all dimensions, and boundary conditions in terms of pressures or fluxes. Moreover, all fractures need a specified aperture, the transverse area of 1D and 0D intersections are then computed by the product of the meeting fractures. The coupling coefficient $K_n$ can also be provided.

### 3.1.2 Discretization

PorePy provides three discretization schemes for the flow equation: The standard two-point flux approximation (TPFA) (Aziz and Settari, 1979) and its extension the multi-point flux approximation (MPFA) (Aavatsmark, 2002) are both well established for subsurface flow. TPFA is the industry standard within petroleum simulations, while MPFA has superior accuracy for anisotropic permeabilities. The mixed-dimensional coupling for TPFA and MPFA follows (Karimi-Fard et al., 2004).

As indicated above, meshing of complex fracture networks tends to result in a high number of cells, which impair computational efficiency. Cell merging into general polyhedrons can partly alleviate the cost, but both TPFA and MPFA run into problems for such grids. PorePy therefore provides discretization by the mixed virtual element method (Beirão da Veiga et al., 2016a, 2016b), which is ideally suited to solve the pressure equation on general cells. For details see (Fumagalli and Keilegavlen, 2017).

## 3.2 Heat transport

The implemented model handles advection from a pre-computed velocity field, for instance obtained from a pressure solve, coupled with a diffusion term.

### 3.2.1 Model and parameters

Transport of a scalar, here denoted $T$ for temperature, is modeled by the advection-diffusion equation

$$(\rho c)_{eff} \frac{\partial T}{\partial t} + (\rho c)_f (\boldsymbol{q} \cdot \nabla T) - \nabla \cdot (\boldsymbol{C}_{eff} \nabla T) = w_T. \tag{3}$$

Here, $\rho$ denotes density, $c$ is heat capacity per volume, and $\boldsymbol{C}$ denotes the thermal conductivity. The subscript $f$ denotes fluid, while subscript $eff$ denotes the porosity-weighted average between fluid and rock properties. The source term $w_T$ denotes the toal heat source, and boundary conditions are either fixed temperatures or effective heat fluxes. These parameters are all specified in a PorePy simulation. The flow field $\boldsymbol{q}$ is computed from the pressure equation (1) - (2). The coupling between dimensions is modeled by conservation of energy.

### 3.2.2 Discretization

In PorePy, the transport equation is discretized using a finite volume approach. The advection term is discretized by a standard upwind scheme, while the conduction term can be handled both by TPFA and MPFA. Both implicit and explicit time stepping schemes are available, as are more advanced options such as Crank-Nicholson and BDF2.

## 3.3 Rock Mechanics

Contrary to fluid flow and transport, the mechanical behavior of the rock is commonly modeled by different equations than in the fracture, e.g. (Aagaard et al., 2013). Accordingly, the elasticity module in PorePy does not couple dynamics between dimensions, but considers the domain of the highest dimension only.

### 3.3.1 Model

To incorporate mechanical deformation, the rock is modeled as a quasi-static linearly elastic medium, governed by the equation

$$\nabla \cdot \left( 2\mu \frac{(\nabla \boldsymbol{d}_M + (\nabla \boldsymbol{d}_M)^T)}{2} + \lambda I \left( \nabla \cdot \boldsymbol{d}_M \right) \right) = \boldsymbol{b}_M.$$

Here, $\mu$ and $\lambda$ are the Lamé parameters (functions to translate from other elastic modulii are provided), $\boldsymbol{d}_M$ is the deformation, and $\boldsymbol{b}_M$ represents body forces. On fracture surfaces, the slip distance is equal to the displacement jump,

$$\boldsymbol{s} = \boldsymbol{d}_F^+ - \boldsymbol{d}_F^-,$$

where $\boldsymbol{d}_F^+$ and $\boldsymbol{d}_F^-$ represent the displacements on the two sides of the fracture. The relation between tractions and deformation on the fracture surfaces is governed by a joint deformation model, such as (Barton et al., 1985).

### 3.3.2 Discretization

Within PorePy the rock mechanics equations are discretized using a cell-centered finite volume approach known as multi-point stress approximation (MPSA) methods (Keilegavlen and Nordbotten, 2017; Nordbotten, 2014). The fractures are linked to the surrounding rock through conditions on internal boundaries corresponding to the fracture surface, with displacements on each side of the fractures, $\boldsymbol{d}_F$, as variables, see (Ucar et al.) for details. The MPSA implementation also covers the extension to poro-elasticity developed in (Nordbotten, 2016), which can be of relevance for certain rock types.

## 3.4 Additional physical relations

In addition to the processes and equations described above, it will often be of interest to include other constitutive relations in the simulation models. Some examples of these are equations of state, non-linear friction laws for fracture surfaces, and relations between fracture sliding and aperture increases (Barton et al., 1985; Olsson and Barton, 2001). These relations are typically application specific, and are therefore not included in PorePy. Their implementation in a simulation model is often relatively simple, exploiting libraries for numerical computations within *Python.*

## 3.5 Linear solvers

Most equations and discretization schemes implemented in PorePy result in a linear system to be solved. For large problems this is computationally highly demanding, and may consume a significant portion of the overall simulation time. PorePy tries to shield the user from this by selecting solvers based on the problem type and size, this setting may easily be overridden if desirable.

## 4. Multi-physics couplings

To increase the number of physical processes that can be captured, as well as ensuring flexibility in the coupling schemes, PorePy is written in a modular style that emphasizes single processes rather than couplings. The simulation of multi-physics problems often requires user-specified coupling schemes. The *Python* scripting environment makes these relatively simple to design. Here, we present two examples of such processes, the first a coupling between the pressure equation and transport based on the subsequent velocity field, while the second considers coupling of flow, linear elasticity in the rock matrix and non-linear shear deformation of fractures. Scripts for both simulations can be accessed on the PorePy repository.

## 4.1 Flow and transport

We consider temperature transport through the three-dimensional domain created by outcrop extrusion, presented in Figure 4. The computational mesh consists of 243352 3D cells, 28113 2D cells and 256 1D cells; the fracture network did not contain any intersections of lines, thus no 0D cells were needed. For this domain, we impose a gradient in pressure and temperature across the domain, and use PorePy to first calculate the pressure field, and then calculate temperature transport by combined advection and diffusion, with advection being based on fluxes derived from the pressure solution. For simplicity, we do not account for compressibility or temperature effects in the pressure equation. The fracture network is not fully connected, forcing the fluid to also flow through the less permeable matrix.

As can be seen from the pressure and temperature profiles depicted in Figure 5, the PorePy simulation model can resolve the interaction between fractures and the host rocks. A code snippet, showing the minimal code necessary to set up a coupled problem for flow and transport, is shown in Figure 6; the full code including parameter setup can be found in the PorePy repository.

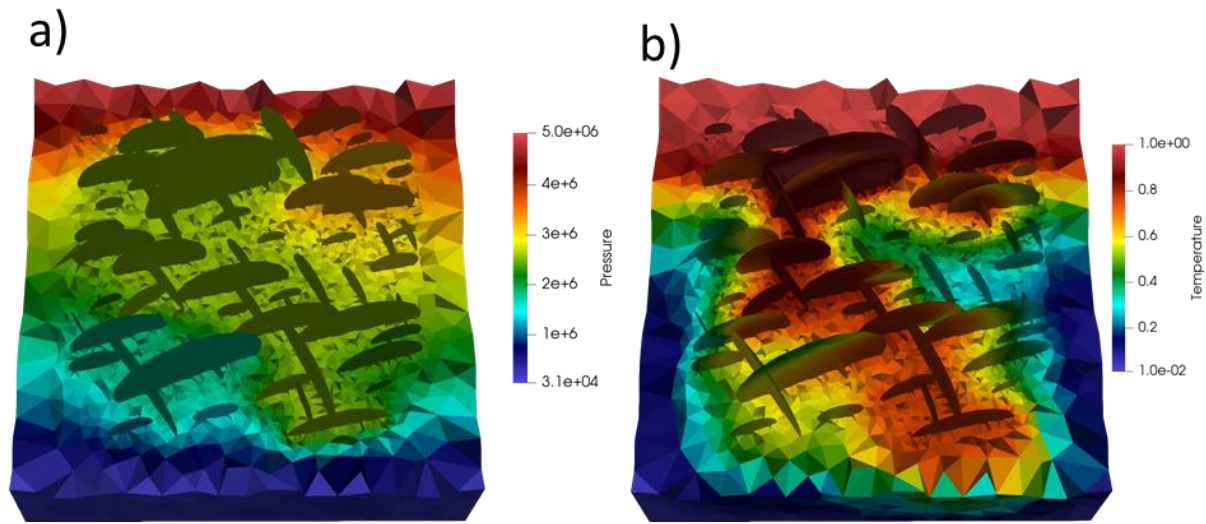

*Figure 5: Solution to a coupled flow and transport problem. The dynamics is driven by a gradient in pressure and temperature from back to front, with no-flow and insulated boundaries on the sides. a) Pressure solution, b) scaled temperature.*

```
# Read fracture definition from file
# The mesh object also contains data and solution
mesh = meshing.simplex_grid(fractures)

# Define pressure solver with default data
pressure_solver = EllipticModel(mesh)
# Solve pressure equation
pressure_solver.solve()
# Save pressure solution to file
pressure_solver.save()

# Derive fluxes from computed pressure field
compute_discharge(mesh)

# Define transport solver, using derived fluxes
transport_solver = ParabolicModel(mesh)
# Solve transport problem
transport_solver.solve()
# Save solution
transport_solver.save()
```

*Figure 6: Minimal code snippet necessary to set up and solve a problem with Darcy flow, and subsequent temperature transport by combined advection and diffusion. The code used to produce Figure 5, including specification of parameters, can be found in the PorePy repository.*

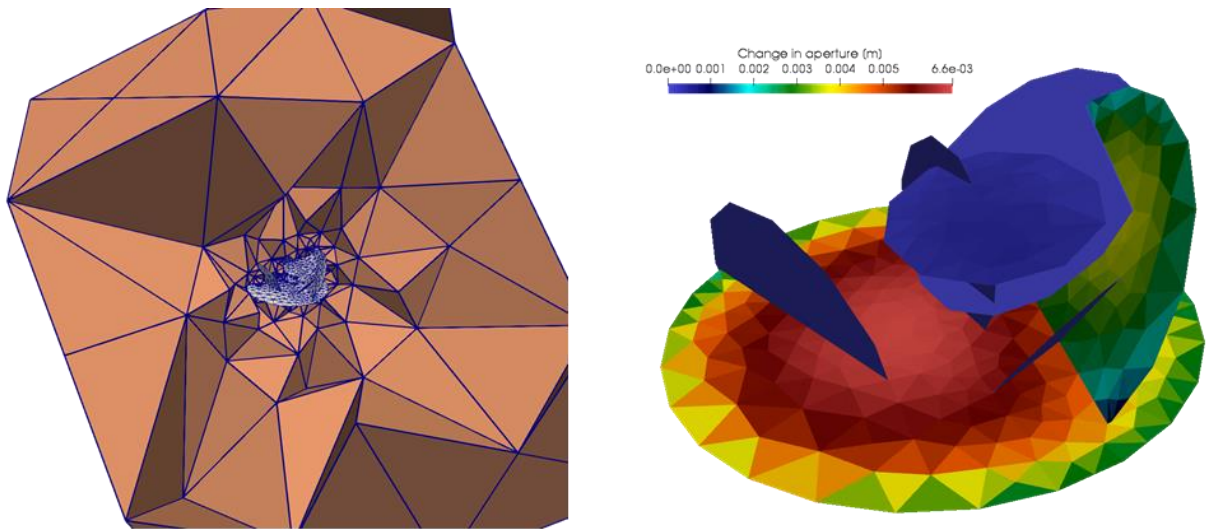

*Figure 7: Slip distance after hydraulic stimulation, with injection in the lowermost fracture. Aperture and permeability changes can be calculated from the slip distance. The magnitude of the changes is determined both by fluid pressure and the fracture orientation relative to anisotropies in the background stress field.*

## 4.2 Low-pressure stimulation of a geothermal reservoir

As a second example of multi-physics couplings, we consider fracture reactivation due to fluid injection in a fracture network. This process is relevant for low-pressure stimulation for enhanced geothermal systems, and entails coupling of fluid flow, rock deformation and a friction law for fracture slip. Slip is triggered as the frictional force is lowered by fluid pressure inside fractures, thus an accurate representation of fracture flow is key for such simulations. Details on the numerical coupling scheme implemented in PorePy can be found in (Ucar et al., 2017a, 2017b).

```python
# Initialize mesh with data
...

# Define the flow, elasticity and friction models
flow_solver = SlightlyCompressibleModel(mesh)
mech_solver = StaticModel(mesh)
friction_solver = FrictionSlipModel(mesh)

# define time iteration loop
while t <= end_time:
    t += time_step

    # Reassemble flow discretization, using updated apertures
    # and injection rates
    flow_solver.reassemble(t)
    # Propagate flow solution
    flow_solver.step()
    # Update pressure state, for use in mechanics solve
    update_pressure(mesh, flow_solver)

    # Equilibrate rock deformation and sliding along fractures
    # Take at least one iteration
    do_slip = True
    while do_slip:
        # Solve linear elasticity, given current state of slip
        mech_solver.solve()
        # Update traction, for use in friction model
        update_traction(mesh, mech_solver)
        # Estimate new slip state, given updated stress field
        # do_slip is False if friction and shear traction
        # is in equilibrium
        do_slip = friction_solver.solve()
        # Update slip state in mech_solver
        update_slip(mesh, friction_solver)

    # Update aperture state, ready for next flow solver
    update_aperture(mesh, friction_solver)
```

*Figure 8: Code snippet indicating the PorePy simulation of hydraulic stimulation. The full simulation script can be found in the PorePy repository.*

We consider fluid injection in a three-dimensional fractured reservoir for a short period, followed by pressure migration through the fracture network and the surrounding rock matrix. Figure 7 depicts the accumulated aperture changes at the end of the simulation. We observe that the aperture changes vary significantly within fracture planes. This shows the utility of the high resolution of the pressure front, but also of the ability to represent local variations in parameters, in this case the aperture and fracture permeability. A code snippet indicating the PorePy code necessary for the simulation is shown in Figure 8.

## 5. Concluding remarks

The open-source framework PorePy enables rapid prototyping for simulation of dynamics in fractured porous media. PorePy gives users easy access to modern simulation tools, including automatic generation of meshes that conform to fractures, and discretization schemes for flow, transport and rock mechanics. A modular software design allows rapid exploration of multi-physics couplings. Two demonstration cases simulating temperature transport and hydroshearing in complex 3D fracture networks using the code were shown.

To access the software, including run scripts for the simulations in section 4, other examples and tutorials, and future enhancements of the code, see www.github.com/pmgbergen/porepy.

## Acknowledgements

The software reported in this paper was financed in part by the Research Council of Norway (grant no 244129, 244035, 267908, 250223).

## Bibliography

Aagaard, B.T., Knepley, M.G., Williams, C.A., 2013. A domain decomposition approach to implementing fault slip in finite-element models of quasi-static and dynamic crustal deformation. J. Geophys. Res. Solid Earth 118, 3059–3079. doi:10.1002/jgrb.50217

Aavatsmark, I., 2002. An introduction to multipoint flux approximations for quadrilateral grids. Comput. Geosci. 6, 405–432. doi:10.1023/A:1021291114475

Ahrens, J., Geveci, B., Law, C., 2005. ParaView: An end-user tool for large data visualization. Elsevier.

Aziz, K., Settari, A., 1979. Petroleum Reservoir Simulation. Applied Science Publishers.

Barton, N., Bandis, S., Bakhtar, K., 1985. Strength, deformation and conductivity coupling of rock joints. Int. J. Rock Mech. Min. Sci. Geomech. Abstr. 22, 121–140. doi:10.1016/0148-9062(85)93227-9

Beirão da Veiga, L., Brezzi, F., Marini, L.D., Russo, A., 2016a. Mixed virtual element methods for general second order elliptic problems on polygonal meshes. ESAIM Math. Model. Numer. Anal. 50, 727–747. doi:10.1051/m2an/2015067

Beirão da Veiga, L., Brezzi, F., Marini, L.D., Russo, A., 2016b. H(div) and H(curl) -conforming virtual element methods. Numer. Math. 133, 303–332. doi:10.1007/s00211-015-0746-1

Berkowitz, B., 2002. Characterizing flow and transport in fractured geological media: A review. Adv. Water Resour. 25, 861–884. doi:10.1016/S0309-1708(02)00042-8

Bisdom, K., Gauthier, B.D.M., Bertotti, G., Hardebol, N.J., 2014. Calibrating discrete fracture-network

models with a carbonate three-dimensional outcrop fracture network: Implications for naturally fractured reservoir modeling. Am. Assoc. Pet. Geol. Bull. 98, 1351–1376. doi:10.1306/02031413060

Boon, W.M., Nordbotten, J.M., Yotov, I., 2016. Robust Discretization of Flow in Fractured Porous Media. arXiv:1601.06977.

Dietrich, P., Helmig, R., Sauter, M., Hötzl, H., Köngeter, J., Teutsch, G. (Eds.), 2005. Flow and Transport in Fractured Porous Media. Springer-Verlag, Berlin/Heidelberg. doi:10.1007/b138453

Fakcharoenphol, P., Xiong, Y., Hu, L., Winterfeld, P.H., Xu, T., Wu, Y.-S., 2013. User's guide of Tough2 EGS-MP: A massively parallel simulator with coupled geomechanics for fluid and heat flow in enhanced geothermal systems.

Flemisch, B., Berre, I., Boon, W., Fumagalli, A., Schwenck, N., Scotti, A., Stefansson, I., Tatomir, A., 2018. Benchmarks for single-phase flow in fractured porous media. Adv. Water Resour. 111, 239–258. doi:10.1016/j.advwatres.2017.10.036

Flemisch, B., Darcis, M., Erbertseder, K., Faigle, B., Lauser, A., Mosthaf, K., Müthing, S., Nuske, P., Tatomir, A., Wolff, M., Helmig, R., 2011. DuMux: DUNE for multi-{phase,component,scale,physics,...} flow and transport in porous media. Adv. Water Resour. 34, 1102–1112. doi:10.1016/j.advwatres.2011.03.007

Fumagalli, A., Keilegavlen, E., 2018. Dual virtual element method for discrete fractures networks. SIAM J. Sci. Comput. Accepted, arXiv:1610.02905v1.

Fumagalli, A., Keilegavlen, E., 2017. Dual virtual element methods for discrete fracture matrix models. arXiv: 1711.01818.

García-Sellés, D., Falivene, O., Arbués, P., Gratacos, O., Tavani, S., Muñoz, J.A., 2011. Supervised identification and reconstruction of near-planar geological surfaces from terrestrial laser scanning. Comput. Geosci. 37, 1584–1594. doi:10.1016/j.cageo.2011.03.007

Geuzaine, C., Remacle, J.-F., 2009. Gmsh: A 3-D finite element mesh generator with built-in pre- and post-processing facilities. Int. J. Numer. Methods Eng. 79, 1309–1331. doi:10.1002/nme.2579

Hardebol, N.J., Bertotti, G., 2013. DigiFract : A software and data model implementation for flexible acquisition and processing of fracture data from outcrops. Comput. Geosci. 54, 326–336. doi:10.1016/j.cageo.2012.10.021

Karimi-Fard, M., Durlofsky, L.J., Aziz, K., 2004. An efficient discrete-fracture model applicable for general-purpose reservoir simulators. SPE J.

Keilegavlen, E., Nordbotten, J.M., 2017. Finite volume methods for elasticity with weak symmetry. Int. J. Numer. Methods Eng. 112, 939–962. doi:10.1002/nme.5538

Kolditz, O., Bauer, S., Bilke, L., Böttcher, N., Delfs, J.O., Fischer, T., Görke, U.J., Kalbacher, T., Kosakowski, G., McDermott, C.I., Park, C.H., Radu, F., Rink, K., Shao, H., Shao, H.B., Sun, F., Sun, Y.Y., Singh, A.K., Taron, J., Walther, M., Wang, W., Watanabe, N., Wu, Y., Xie, M., Xu, W., Zehner, B., 2012. OpenGeoSys: an open-source initiative for numerical simulation of thermo-hydro-mechanical/chemical (THM/C) processes in porous media. Environ. Earth Sci. 67, 589–599. doi:10.1007/s12665-012-1546-x

Lee, S.H., Lough, M.F., Jensen, C.L., 2001. Hierarchical modeling of flow in naturally fractured formations with multiple length scales. Water Resour. Res. 37, 443–455.

Lie, K., Krogstad, S., Ligaarden, I.S., Natvig, J.R., Nilsen, H.M., Skaflestad, B., 2012. Open-source MATLAB implementation of consistent discretisations on complex grids. Comput. Geosci. 16, 297–322. doi:10.1007/s10596-011-9244-4

Martin, V., Jaffré, J., Roberts, J.E., 2005. Modeling Fractures and Barriers as Interfaces for Flow in Porous Media. SIAM J. Sci. Comput. 26, 1667–1691. doi:10.1137/S1064827503429363

Matthai, S.K., Geiger, S., Roberts, S.G., Paluszny, a., Belayneh, M., Burri, a., Mezentsev, a., Lu, H., Coumou, D., Driesner, T., Heinrich, C. a., 2007. Numerical simulation of multi-phase fluid flow in structurally complex reservoirs. Geol. Soc. London, Spec. Publ. 292, 405–429. doi:10.1144/SP292.22

Nordbotten, J.M., 2016. Stable cell-centered finite volume discretization for Biot equations. SIAM J. Numer. Anal. 54, 942–968. doi:10.1137/15M1014280

Nordbotten, J.M., 2014. Cell-centered finite volume discretizations for deformable porous media. Int. J. Numer. Methods Eng. 100, 399–418. doi:10.1002/nme.4734

Olsson, R., Barton, N., 2001. An improved model for hydromechanical coupling during shearing of rock joints. Int. J. Rock Mech. Min. Sci. 38, 317–329. doi:10.1016/S1365-1609(00)00079-4

Open Porous Media Team, 2017. Open porous media flow documentation manual [WWW Document].

Paluszny, A., Matthäi, S.K., Hohmeyer, M., 2007. Hybrid finite element-finite volume discretization of complex geologic structures and a new simulation workflow demonstrated on fractured rocks. Geofluids 7, 186–208. doi:10.1111/j.1468-8123.2007.00180.x

Peacock, D.C.P., Nixon, C.W., Rotevatn, A., Sanderson, D.J., Zuluaga, L.F., 2017. Interacting faults. J. Struct. Geol. 97, 1–22. doi:10.1016/j.jsg.2017.02.008

Preuss, K., 1991. TOUGH2 a General-Purpose Numerical Simulator for Multiphase Fluid and Heat Flow.

Rotevatn, A., Buckley, S.J., Howell, J.A., Fossen, H., 2009. Overlapping faults and their effect on fluid flow in different reservoir types: A LIDAR-based outcrop modeling and flow simulation study. Am. Assoc. Pet. Geol. Bull. 93, 407–427. doi:10.1306/09300807092

Sanderson, D.J., Nixon, C.W., 2015. The use of topology in fracture network characterization. J. Struct. Geol. 72, 55–66. doi:10.1016/j.jsg.2015.01.005

Ucar, E., Berre, I., Keilegavlen, E., 2017a. Three-dimensional numerical modeling of shear stimulation of naturally fractured rock formations. arXiv:1709.01847.

Ucar, E., Berre, I., Keilegavlen, E., 2017b. Postinjection normal closure of fractures as a mechanism for induced seismicity. Geophys. Res. Lett. 44, 9598–9606. doi:10.1002/2017GL074282

Ucar, E., Keilegavlen, E., Berre, I., Nordbotten, J.M.,. Finite-volume discretization for the deformation of fractured media. arXiv: 1612.06594.